# Estimating Earthquake Early Warning Effectiveness via Blind Zone Sizes: A Case Study of the Planned Seismic Network in Chinese Mainland

**Jiawei Li** (Corresponding author; lijw@cea-igp.ac.cn; lijw3@sustech.edu.cn)

Institute of Risk Analysis, Prediction and Management (Risks-X), Academy for Advanced Interdisciplinary Studies, Southern University of Science and Technology (SUSTech), Shenzhen, China, 518055

https://orcid.org/0000-0003-1226-8162

**Didier Sornette**

Institute of Risk Analysis, Prediction and Management (Risks-X), Academy for Advanced Interdisciplinary Studies, Southern University of Science and Technology (SUSTech), Shenzhen, China, 518055

**Yu Feng**

Institute of Risk Analysis, Prediction and Management (Risks-X), Academy for Advanced Interdisciplinary Studies, Southern University of Science and Technology (SUSTech), Shenzhen, China, 518055

Norwegian Geotechnical Institute, Oslo, Norway, N-0484 (now at)

https://orcid.org/0000-0003-3139-8931

**Abstract** The China Earthquake Administration (CEA) has launched an ambitious nationwide earthquake early warning (EEW) system project currently under development, which will include approximately 15,000 seismic stations and be the largest EEW system in the world. The new EEW system is planned to go online at the end of 2023. In approximately 50%, 30% and 20% of Chinese mainland, the inter-station distance will soon be smaller than 50 km, 25 km and 15 km, respectively. The expected effectiveness of this EEW system can be quantified via the metric determined from the radius of the blind zone, which refers to the area near the epicenter where there is insufficient time to issue a warning before the arrival of strong *S*- and surface waves. This study uses a theoretical network-based method together with Monte Carlo simulation to obtain the spatial distribution of the blind zone radii and their associated uncertainties for the new seismic network based on its configuration. We find that the densified new seismic network is expected to have excellent EEW performance as the area covered by small blind zones with radius ≤ 30 km increases dramatically from approximately 2% to 22%, that is by 2.4 million km$^2$ inside Chinese mainland. We also find that every 1,000,000 RMB (≈146,000 USD) invested to densify the planned network will lead to an areal increase of 3,000 km$^2$ of small blind zones. Continuing to increase the density of stations in some key regions with blind zone radii ranging from 15 to 40 km is still necessary to control the unexpected expansion of blind zones due to possible (and common) stations failure. Our work provides insights into the expected performance of the upcoming EEW network in Chinese mainland, and our proposed evaluation approach is broadly applicable for predicting the performance of EEW systems during their planning, design, and implementation stages.

## ACKNOWLEDGEMENTS

The authors would like to thank Prof. Zhongliang Wu at the Institute of Earthquake Forecasting of China Earthquake Administration (CEA) for discussion about blind zone, Dr. Chen Yang and Dr. Yang Zang at the China Earthquake Networks Center for providing information on the stations. This project is supported by the Guangdong Basic and Applied Basic Research Foundation under Grant No. 2020A1515110844, the Special Fund of the Institute of Geophysics, China Earthquake Administration under Grant No. DQJB22Z01-10, and the National Natural Science Foundation of China under Grant No. U2039202 and U2039207.

## INTRODUCTION

The concept of the blind zone, also referred to as the no-alert zone or late-alert zone, and previously termed as the blind zone in earlier literature, is a crucial aspect of the Earthquake Early Warning (EEW) system. It designates the geographic vicinity surrounding the epicenter where there is insufficient time for issuing warnings before the onset of intense *S*- and surface seismic waves (Kuyuk and Allen 2013). The radius of the blind zone assumes a significant role in evaluating the effectiveness of EEW systems. This metric directly signifies the system's capability to provide prompt alerts (with more positive lead time or available warning time) to regions at risk, thereby mitigating potential damage and safeguarding lives (Picozzi et al. 2015; Pan et al. 2019; Wu et al. 2019; Cremen et al. 2022). A smaller radius denotes a more efficient and robust EEW system, ensuring that vulnerable regions receive advance warnings of impending destructive seismic waves. The densification of a seismic network reduces blind zones of EEW system (Kuyuk and Allen 2013). As the time when the system alert triggers relative to the origin time, system alert trigger latency $T_{\mathrm{tri}}$ has been considered to be the reason for the existence of blind zone for shallow earthquakes. $T_{\mathrm{tri}}$ is determined by many factors, e.g., network density, focal depth, seismic velocity structure of the crust, speed of data transmission and system processing (Kuyuk and Allen 2013; Yang et al. 2015; Picozzi et al. 2015; Guo et al. 2016; Li and Wu 2016; Festa et al. 2018).

The blind zone radius $r$ is the simplest and one of the best proxies of the overall performance of an EEW system. Although it may not be obvious whether an alert within a blind zone can be acted upon in a timely or useful manner, an EEW system can still be useful for the sites outside the blind zone near the fault rupture, which is especially relevant for earthquakes with hundred kilometers ruptures (Böse et al. 2018, 2021; Ma et al. 2012; Li et al. 2018, 2020). It is a geometric fact that a blind zone with a radius of 10 km to 20 km is inevitable for a shallow earthquake regardless of the network density (Kuyuk and Allen 2013; Wald 2020). Hence, the optimization of an EEW system mainly targets to minimize the size of the blind zone, and in this study, we evaluate how the areas with small blind zones are extended with the densification of the new nationwide station.

Inspired by the developments of the EEW systems, and also the benefit from them, in other regions worldwide (Allen et al. 2009; Satriano et al. 2011a; Strauss and Allen 2016; Allen and Melgar 2019; Allen and Stogaitis 2022), Chinese mainland has been continuously developing and evolving its EEW systems at the city/infrastructure and regional/provincial scales in order to improve their capability for earthquake risk mitigation in the past three decades (e.g., Li et al. 2004; Peng et al. 2011; Li 2014; Zhang et al. 2016; Peng et al. 2020, 2021, 2022). To further extend the scale and applications of the existing seismic network, the National System for Fast Report of Intensities and Earthquake Early Warning Project of China launched by the China Earthquake Administration (CEA) has started a nationwide network construction project with a total investment of approximately 1.87 billion RMB (290 million USD; see Data and Resources). Through this project, the number of seismic stations will increase from approximately 2,000 to 15,000 in the near future, with data transmission for all stations planned to have very little latency (Jiang and Liu 2016). The theoretical network-based approach proposed by Kuyuk and Allen (2013) is suitable for predicting the performance of the planned EEW system, as it relies solely on the spatial distribution of stations. Following a similar approach, Pan et al. (2019) performed a preliminary analysis of system alert trigger latency for the existing prototype and future EEW systems in Gansu province, China, according to their configurations. Their projections indicated significant improvements in the operational efficiency of the upcoming EEW system in the southeastern region of Gansu. The anticipated system alert trigger latency ranged between 7 to 12 s, corresponding to a blind zone radius spanning from 20 to 40 km.

Unlike the real single case-based analysis (e.g., Satriano et al. 2011b; Pazos et al. 2015; Li et al. 2018; Hsu et al. 2018; Peng et al. 2021, 2022), the theoretical network-based method requires several hypotheses (Kuyuk and Allen 2013; Li and Wu 2016). The minimum number of trigger stations is a critical one among them. The majority of the EEW systems worldwide adopt two to four stations as the minimum number to trigger. However, due to stations near the epicenter failing to work (Bormann 2002) or underestimating ground-shaking hazards (Ma et al. 2012) in the real world, the EEW system has to require a larger number of stations to trigger, which implies increasing the inter-station distance and therefore expanding the blind zone. Consequently, Li and Wu (2016) proposed that there is a possible "soft" blind zone around the original ("hard") blind zone when taking account of the potential stations closest to the epicenter failure, and argued that the "soft" blind zone could be easily controlled by simply densifying the seismic network. Although an unexpected expansion of the blind zone ("soft" blind zone) is common and challenges the success of an EEW system, so far there is no study that quantitatively investigates how the EEW effectiveness is affected by the "soft" blind zone. Another important issue with the network-based method is that the estimated blind zone radii are subject to parameter uncertainty, which has largely been ignored in the existing applications.

In this study, we predict the EEW performance estimated by the blind zone radii for the planned seismic network and assess their uncertainties in terms of probability distributions by employing the Monte Carlo simulation method. We begin with a brief introduction to the new EEW system in Chinese mainland. Then, we predict the improvement of the planned network compared with the existing network from the perspective of the spatial distribution of blind zones and their uncertainties. Finally, the significance of controlling the "soft" blind zones is discussed quantitatively. Our work offers insights into the anticipated performance of the forthcoming EEW network in Chinese mainland. Furthermore, the proposed evaluation approach can be applied beyond our specific case, for instance for predicting the performance of any EEW system during its planning, design and implementation stages.

## DATA

In order to estimate the blind zone radii, we generate random seismogenic depths $H$ using Monte Carlo simulations, based on the probability distribution of depths derived from observed seismicity. From January 1, 1970, to June 26, 2022, approximately 1,586,600 earthquakes were observed in China and its adjacent regions, and 91,891 events of them have $M_L \geq 3.0$ as shown in Figure S1. Of this catalogue, the 2008 $M_S$ 8.0 Wenchuan earthquake is the largest and most catastrophic event, and resulted in about 70,000 fatalities and 20,000 missing people.

The new seismic network in Chinese mainland consists of a total of around 15,200 stations (Figure S2): approximately 2,000 datum stations equipped with broadband seismometers and accelerometers, 3,200 basic stations equipped with only accelerometers and 10,000 ordinary stations equipped with low-cost micro-electro-mechanical system (MEMS) intensity sensors. For the purpose of this study, we define the inside of Chinese mainland as the region within 100 km outside of the boundaries of Chinese mainland and Hainan Island (approximately 11.5 million km$^2$). The inter-station distance at a given site (assuming a grid of 0.1° × 0.1° resolution) is defined as the average distance from this site to its three closest stations (Kuyuk and Allen 2013). In approximately 80% of the area inside Chinese mainland, the inter-station distance of the densified new network is less than 100 km, 50% less than 50 km, and 30% less than 25 km (Figure S3). The new network is expected to be officially online at the end of 2023 (Wenhui Huang, Chaoyong Peng, and Peng Jiang, written comm. 2023). The new network will be the main backbone of the largest EEW system in the world. In the future, EEW services will be provided in five key regions (North China, the central China north-south seismic belt, the southeast coast of China, the middle section of the Tianshan Mountains in Xinjiang, and Lhasa in Tibet; Figure S1). The seismic intensity will be reported rapidly within minutes from the origin time for the entire Chinese mainland, and the monitoring capability in Chinese mainland will be significantly improved (Li et al. 2023).

To compare with the planned EEW system, we envisage a virtual EEW system by combining the existing high-gain broadband seismic and strong motion networks with a total of approximately 950 and 1,100 stations, respectively (Figure S2). It's important to note that, among these stations, only high-gain broadband seismic stations have real-time data transmission capabilities; strong motion stations lack this feature. It is worth emphasizing that neither high-gain broadband seismic stations nor strong motion stations were originally designed for EEW purposes. The virtual EEW system is formulated within the existing seismic network under the assumption that all stations possess real-time data transmission capabilities. Only in 55% of inside Chinese mainland, the inter-station distances of the existing network are less than 100 km, 24% are less than 50 km, and 7% are less than 25 km (Figure S3).

The majority of stations from the existing high-gain broadband seismic and strong motion networks will undergo upgrades to incorporate real-time data transmission capabilities and will be integrated into the planned EEW system. Compared with the existing network, the area with inter-station distances smaller than 15 km that is covered by the new network expands by a factor of approximately 10 to reach 17.3% from 1.7% (Figure S3). The areas with inter-station distances smaller than 100 km, 50 km and 25 km will be expanded by factors of 1.4, 2.2 and 4.4 in the near future, respectively. The density of the new network is thus significantly improved.

## EARTHQUAKE EARLY WARNING PERFORMENCE ESTIMATION

### *Blind Zone Radius*

Considering the relatively low observational quality associated with the most ordinary stations equipped with MEMS sensors, we choose conservatively for this study a minimum number of three trigger stations. Hypothesizing a homogeneous Earth crust model, the *P*-wave travel time $t_P$ at the third closest station with epicenter distance $\Delta$ from a point source with depth $H$ is given by:

$$t_{\mathrm{P}} = \frac{\sqrt{\Delta^2 + H^2}}{V_{\mathrm{P}}} \tag{1}$$

where $V_P$ is the *P*-wave velocity. Then, with the system latency $T$ for processing and data transmission, the horizontal distance $r$ of *S*-wave propagation is

$$r = \sqrt{(t_{\mathrm{P}} + T)^2 V_{\mathrm{S}}^2 - H^2} \tag{2}$$

where $V_S$ is the *S*-wave velocity. This means that the *S*-wave has already arrived before the alarm is released at distances smaller than *r* from the earthquake epicenter. The disk with radius *r* is called the blind zone in the EEW system.

Equation (2) indicates that the size of the blind zone is determined by the system alert trigger latency $T_{tri}$, as well as $V_S$ and *H*. $T_{tri}$ is comprised of $t_P$ and *T*, with *Δ*, *H* and $V_P$ determining $t_P$. To investigate the impact of five factors (*Δ*, *H*, $V_P$, $V_S$, and *T*) on the blind zone size *r*, we perform an one-at-a-time sensitivity analysis of *r* by sequentially varying each parameter while maintaining the other four fixed at constant values of $\Delta$ = 30 km, *H* = 5 km, $V_P$ = 6 km/s, $V_S$ = 3.5 km/s, and *T* = 4 s. Figure S4 shows that *r* has an inverse correlation only with *H* and $V_S$. The source depth *H* contributes significantly to the appearance of a zero blind zone, which occurs when $T_{tri}^2 V_S^2 - H^2 < 0$.

The impact of each parameter on *r* was analyzed quantitatively by measuring the change resulting from a one unit shift, as shown by the slope on the right axis in Figure S4. The results show that $V_S$ and *Δ* have the greatest and least impact on *r*, leading to a change of approximately 9.1 to 9.2 km and 0.1 to 0.6 km, respectively. The other three parameters *H*, $V_P$ and *T* have a similar impact effect on *r*, and respectively cause a change of -5 to 0 km, -5 to -2 km and 3.5 km. In real-world scenarios, the practical amplitude variability of each parameter significantly differs from one unit scale (i.e., 1 km, 1 km/s or 1 s). Specifically, we empirically observed changes generally on the order of magnitude of $10^1$ km for *Δ* (Figure S3), $10^0$ to $10^1$ km for *H* (Figure S1), $10^{-1}$ km/s for $V_P$ (e.g., $V_P$ within a depth of 60 km shown by Han et al. (2022)), $10^{-1}$ km/s for $V_S$ (e.g., $V_S$ within a depth of 60 km shown by Han et al. (2022)), and $10^{-1}$ to $10^0$ s for *T* (e.g., Yang et al. 2015). Considering these changes, we find that, in the real-world, *Δ* and *H* have the greatest impact on *r*, resulting in a magnitude change of *r* by $10^1$ km and $-10^1$ km, respectively. Meanwhile, $V_P$, $V_S$, and *T* result in a magnitude change of *r* by $10^{-1}$ to $10^0$ km.

### *Estimating r Using Monte Carlo Random Sampling*

Monte Carlo random sampling is a statistical technique used to generate random samples from a probability distribution (Fishman 1996; Kalos and Whitlock 2008; Rubinstein and Kroese 2017). In this study, we employ this approach as a novelty to estimate $r$ for each 0.1° × 0.1° grid of spatial resolution. Firstly, we generate 50,000 random samples for each of the four parameters ($H$, $V_P$, $V_S$, and $T$) based on given probability density functions (PDFs) and assume a minimum number of three trigger stations to determine $\Delta$. Then, we use these samples along with Equations (1) and (2) to obtain 50,000 $r$ estimates. For each grid, we derive statistical parameters such as the mean and standard deviation based on these estimates.

The observed seismicity, depicted in Figure S1 for events with $M_L \geq 3.0$, is utilized to construct the frequency-depth distribution, without taking magnitude completeness or cut-off magnitude into account. If the number of seismic events within 50 km of each grid exceeds 50, $H$ is generated randomly based on its PDF, which is normalized from the frequency-depth distribution. Otherwise, $H$ is taken as the average depth. When no earthquakes are reported around the grid, $H$ is set to the default value of 5 km. Previous studies have shown that $V_P$ (Sun and Toksöz 2006; Teng et al. 2013; Wei et al. 2016; Han et al. 2022) and $V_S$ (Li et al. 2013; Bao et al. 2015; Shen et al. 2016; Tao et al. 2018; Xin et al. 2019; Han et al. 2022) of the crust within a depth of 60 km in Chinese mainland exhibit slightly larger values in the east than in the west, with a boundary at approximately 105°E. Therefore, we assigned $V_P$ and $V_S$ in the eastern region (> 105°E) as random numbers following the uniform distributions $U(5.5, 7.5)$ and $U(3.7, 4.4)$, respectively, while in the western region (≤ 105°E), the distributions are $U(4.5, 6.5)$ and $U(3.2, 3.9)$, respectively. The boundaries of these uniform distributions for $V_P$ and $V_S$ are derived from findings within a depth of 60 km presented by Han et al. (2022). $T$ comprises the early data window (typically 3 s) after the $P$-wave onsite and the time for processing and transmitting data. For $T$, we generate random samples assuming a normal distribution $N(4, 0.5^2)$ for the planned network and $N(5, 0.5^2)$ for the existing network, respectively (Wenhui Huang, written comm. 2023). For information about the current system trigger criteria of the planned network that determine $T$, we refer to the Discussion.

After performing Monte Carlo random sampling, the 50,000 samples of $r$ on each grid exhibit a frequency distribution that resembles a normal distribution. As a result, we will use the mean and standard deviation of these samples to describe the spatial distribution of blind zones.

***Map of Blind Zone Radius***

Figures 1a and 2a display the spatial distribution of the mean estimated $r$ for the existing and planned networks, respectively. In the existing network, areas with mean $r \leq 30$ km are mainly limited to major cities such as the Beijing-Tianjin Capital Circle and the Pearl River Delta. However, the planned network is expected to achieve almost complete coverage of areas with mean $r \leq 30$ km in North China, the southeast coast, the central China north-south seismic belt, and the middle section of the Tianshan Mountains in Xinjiang. This represents a significant improvement, with coverage increasing by a factor of 13 from 1.7% to reach 22% in the near future. The binned distribution of mean estimated $r$ in Figures 2a indicates that about half of the region inside Chinese mainland will have a mean $r \leq 50$ km. In 15% and 38% of the area inside Chinese mainland, the mean $r$ of the planned network is expected to be smaller than 25 km and 40 km, respectively, which is of great significance for the successful functioning of the EEW system. Figure S5 illustrates the map of relative changes in the mean $r$ inside Chinese mainland between the existing and planned networks, as presented in Figures 2a and 2b. Positive values denote a relative reduction in the mean $r$ of the planned network compared to the existing network.

Figures 1b and 2b display the spatial distribution of the standard deviation of estimated $r$ for the existing and planned networks, respectively. The maps show that there is a strong spatial correlation between the standard deviation of $r$ with station density. Overall, areas where the standard deviation of $r$ is less than 5 km and 10 km will cover 46% and 77% of the areas inside Chinese mainland for the future network, respectively. For the existing network, these proportions are 16% and 51%, respectively. This implies that densification of the networks will effectively constrain the potential variability in the size of the blind zones.

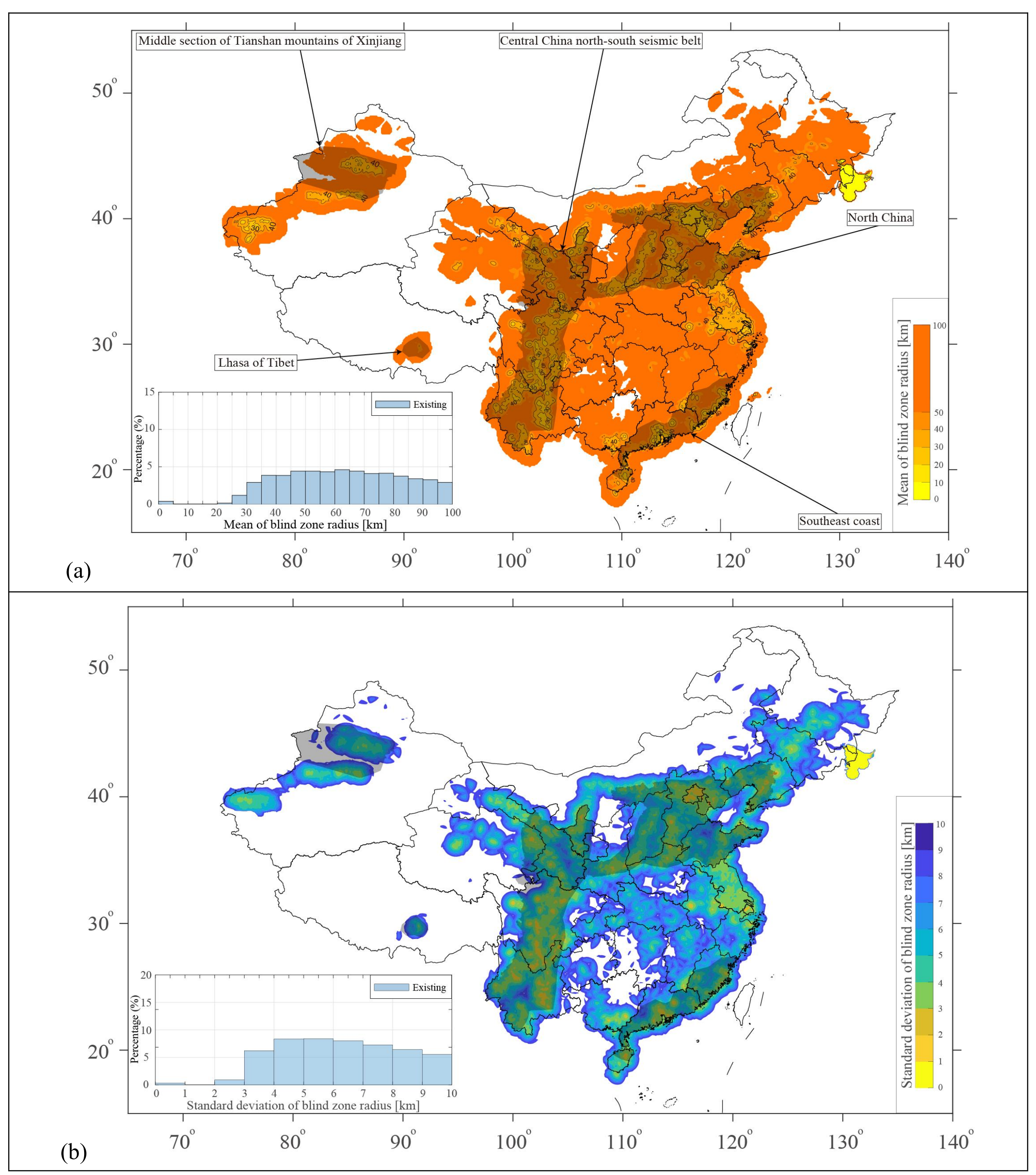

**Figure 1.** Maps showing the (a) mean and (b) standard deviation of the estimated blind zone radii for the existing network, based on a minimum number of three trigger stations and a grid of 0.1° × 0.1° resolution. Areas where the mean and standard deviation of blind zone radii surpass 100 km and 10 km, respectively, are displayed in white. The probability density function (PDF) of the blind zone radius has been constructed using Monte Carlo random sampling, with 50,000 iterations. Further details on the creation of the PDF is available in the text. The histograms in the bottom left depict the binned distribution of mean and standard deviation of the blind zone radii inside Chinese mainland. For this study, we define the "inside" of Chinese mainland as the region within 100 km outside of Chinese mainland's boundaries. The darker areas represent the five key earthquake early warning (EEW) regions that were developed by the National System for Fast Report of Intensities and Earthquake Early Warning Project of China and are expected to be operational after 2023.

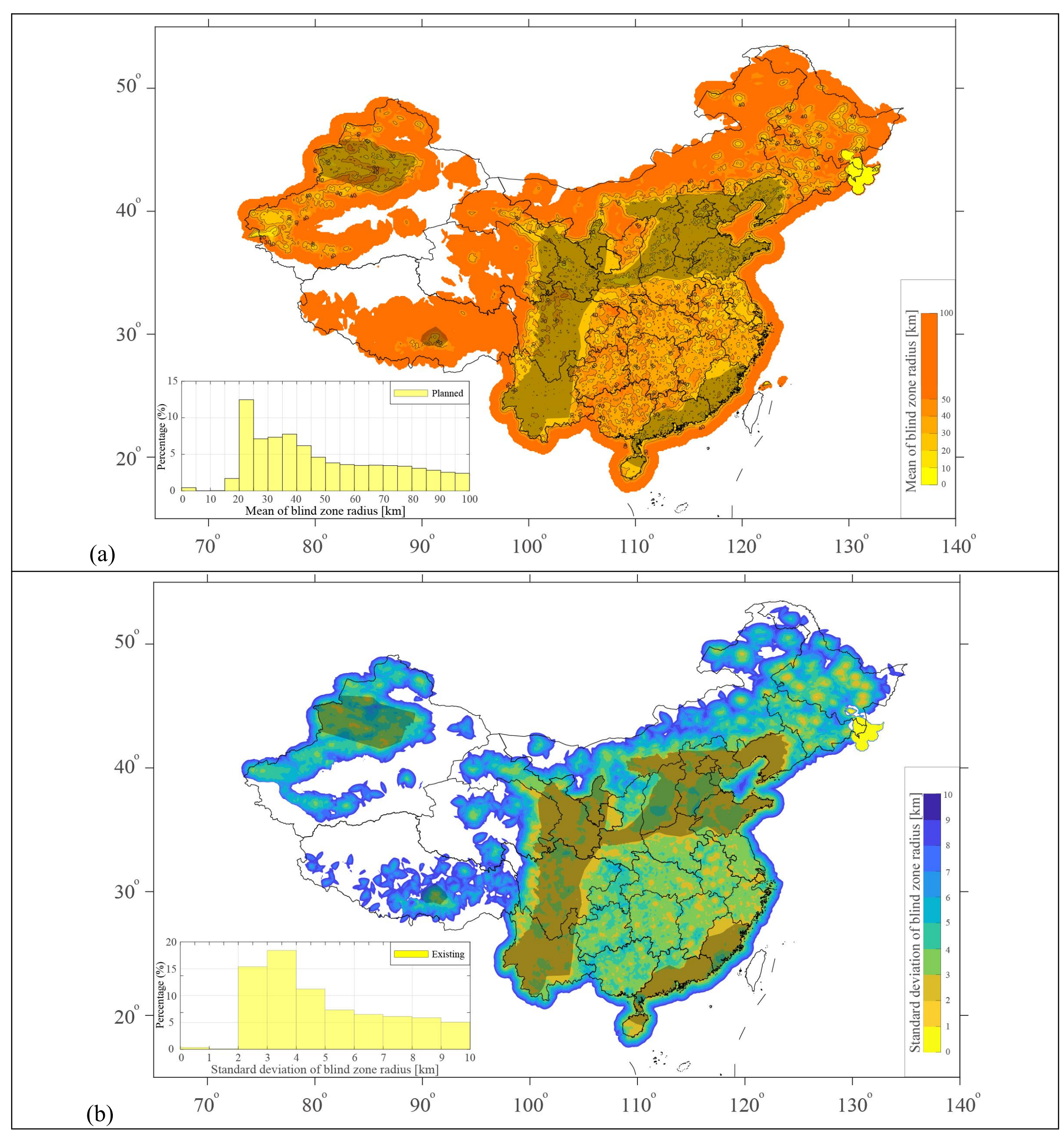

**Figure 2.** Maps showing the (a) mean and (b) standard deviation of the estimated blind zone radii for the planned network. Details are the same as Figure 1.

It should be noted that, for both the existing and planned networks, there are large areas with a mean and standard deviation of zero in Northeast China (bordering Russia and North Korea). This is mainly due to the fact that the focal depth of earthquakes observed in this region is generally greater than 400 km. The Western Pacific Plate subducts beneath the Eurasian Plate and penetrates about 600 km deep into the mainland of Northeast China through the Japan Trench, forming the only deep-focal seismic belt in China (Zhang and Tang 1983; Gudmundsson and Sambridge 1998; Sun and He 2004).

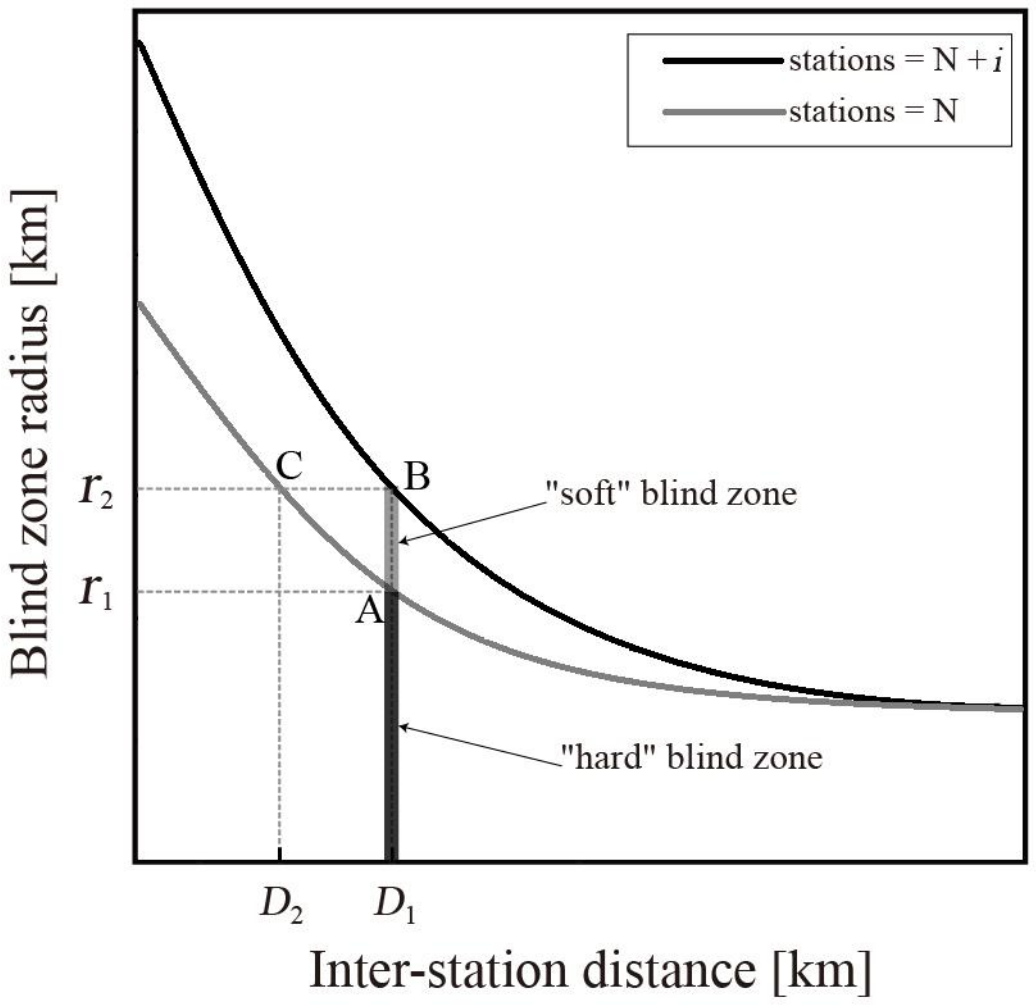

**Figure 3.** The relationship between the blind zone radius ($r$) and inter-station distance ($D$) for various numbers of trigger stations. The $x$-axis represents decreasing inter-station distances from left to right ($D_1 < D_2$) corresponding to an increasing network density, and the $y$-axis increases from bottom to top ($r_1 < r_2$). Modified from Kuyuk and Allen (2013), Figure 5, with simulations assuming earthquakes at a depth of 8 km.

## "SOFT" BLIND ZONE AND ITS CONTROL

Kuyuk and Allen (2013) used a series of evenly distributed modeled shallow point-sources at a depth of 8 km and stations to determine a semi-quantitative empirical curve that describes the relationship between the minimum number of trigger stations, the inter-station distance $D$ and $r$. Figure 3 ($D_1 < D_2$; $r_1 < r_2$) shows that $r$ decreases dramatically with the decrease of inter-station distances when the network density of seismic stations is small. However, for a very dense network, $r$ converges to a plateau, which is approximately the same for different minimum number of trigger stations, ranging from 10 to 20 km depending on different conditions (Kuyuk and Allen 2013).

Figure 3 can also be used to illustrate the equivalence principle between the minimum number of trigger stations and network density from the perspective of blind zones. Assuming a seismic network with an inter-station distance of $D_1$, the minimum number of $N$ trigger stations corresponds to a blind zone of radius $r_1$, as shown in point A of Figure 3. This ideal blind zone with radius $r_1$ is called the theoretical "hard" blind zone by Li et al. (2016). However, in the real world, some $i$ stations near the epicenter may fail to work properly (Bormann 2002) or may not be triggered due to an underestimation of ground-shaking (e.g., YZP station during the 2008 $M_S$ 8.0 Wenchuan earthquake; Ma et al. 2012). Therefore, the EEW system has to adopt $i$ more stations to be triggered, which corresponds to point B with a blind zone radius $r_2$ (where $r_1 < r_2$) in Figure 3. This unexpected and larger blind zone with radius $r_2$ is referred to as the "soft" blind zone by Li et al. (2016). From Figure 3, the size of this "soft" blind zone is equivalent to the size of the "hard" blind zone of a seismic network adopting a minimum number $N$ of trigger stations with inter-station distance of $D_2$ (where $D_1 < D_2$), as shown in point C.

To investigate the performance of the planned EEW system of China (assuming $N$ = 3, same as Figure 2) affected by the "soft" blind zone, we show the cumulative distribution of the mean and standard deviation of estimated $r$ inside Chinese mainland for $i$ = 1 (corresponding to a minimum number of four trigger stations) and $i$ = 2 (corresponding to a minimum number of five trigger stations) potential station failures, respectively (Figure 4). The map showing the mean and standard deviation of $r$ for $N + i = 4$ and 5 can be found in Figures S6 and S7, respectively. The areas with mean $r \leq 20$ km, 30 km, and 40 km account for 1%, 19%, and 32% of the total area inside Chinese mainland for $N + i = 4$, and for 0.6%, 17%, and 27% for $N + i = 5$, respectively. Compared with Figure 2a (without station failures), the area with mean $r \leq 30$ km is reduced by factors of 0.85 and 0.75 for $N + i = 4$ and 5, respectively. The proportion of areas with mean $r \leq 20$ km and 40 km decreases by factors of 0.45 and 0.84 for $N + i = 4$, and 0.29 and 0.72 for $N + i = 5$, respectively (Figure 4a). Compared with Figure 2b, the area with the standard deviation of $r$ smaller than 5 km is reduced by factors of 0.91 and 0.82 for $N + i = 4$ (from 45.5% to 41.25%) and 5 (from 45.5% to 37.47%), respectively (Figure 4b). Overall, continuing to densify the seismic network in some regions where $r$ ranges from 15 to 40 km is still recommended to control the "soft" blind zones and minimize blind zones.

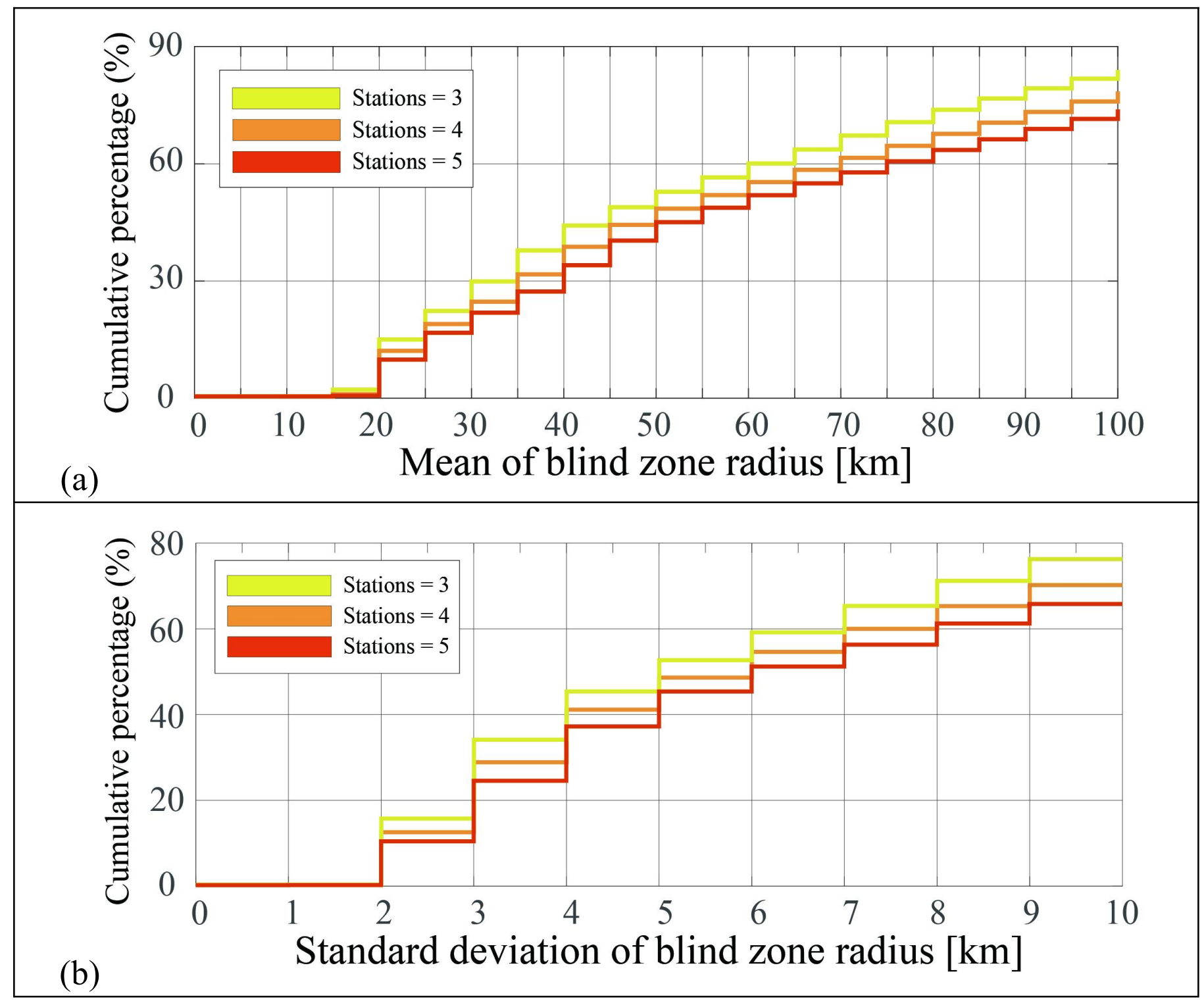

**Figure 4.** The cumulative distribution of the (a) mean and (b) standard deviation of the blind zone radii inside Chinese mainland for the cases of three, four, and five trigger stations, respectively.

## EARTHQUAKE EARLY WARNING EFFECTIVENESS IN FIVE KEY EEW REGIONS

The CEA aims to establish five key EEW regions through the implementation of the National System for Fast Report of Intensities and Earthquake Early Warning Project. The five regions shown in Figure 1 include North China (with an area of 712,747 km$^2$, a population of 316 million, accounting for 23.6% of the national population, and a GDP of 13.05 trillion yuan, accounting for 20.51% of the national GDP), the central China north-south seismic belt (850,905 km$^2$, a population of 109 million, accounting for 8.2% of the national population, and a GDP of 5.15 trillion yuan, accounting for 8.09%), the southeast coast of China (236,502 $km^2$, a population of 108 million, accounting for 8.07% of the national population, and a GDP of 8.12 trillion yuan, accounting for 12.76%), the middle section of the Tianshan Mountains in Xinjiang (235,979 $km^2$, a population of 9 million, accounting for 0.68% of the national population, and a GDP of 0.47 trillion yuan, accounting for 0.74%), and Lhasa in Tibet (30,438 $km^2$). Collectively, these regions account for approximately 40% of the national population and the national GDP, and have experienced earthquakes with $M \geq 6$ in the past (Figure S1). According to the fifth-generation hazard zoning map in China (see Data and Resources), the projected peak ground accelerations (PGA) with an exceedance probability of 10% in the next 50 year are expected to reach 0.1 g across almost the entire area of the five regions, with some parts even reaching 0.3 g. These data were collected up to 2016 and are from the Feasibility Report of the National System for Fast Report of Intensities and Earthquake Early Warning Project of China, which was released within the CEA (courtesy of Changsheng Jiang for providing this report).

Figures 5 and 6 display the cumulative distribution of the mean and standard deviation of the estimated $r$ for the planned network in the five regions, as spatially presented in Figure 2. We find that the areas with mean $r \leq 30$ km in North China and the southeast coast account for approximately 98% of the regions. Areas with mean $r$ smaller than 20 km even account for 14% of the regions. The standard deviation of $r$ in the two regions is less than 5 km, indicating small variability of $r$ due to the high density of the network. The central China north-south seismic belt ranks third, with mean $r \leq 30$ km and 20 km accounting for 91% and 5%, respectively, and a standard deviation smaller than 5 km across the entire region. However, the middle section of the Tianshan Mountains in Xinjiang and Lhasa in Tibet exhibit slightly weaker EEW performance. In the middle section of the Tianshan Mountains in Xinjiang, the coverage of mean $r \leq 30$ km and 20 km is 63% and 6%, respectively, with a standard deviation below 5 km in 87% of the region. Lhasa in Tibet has a relatively poor EEW ability with only 3% coverage of mean $r \leq 50$ km, and the coverage of the 5 km and 10 km standard deviations is 2% and 87%, respectively, making it difficult to achieve effective EEW.

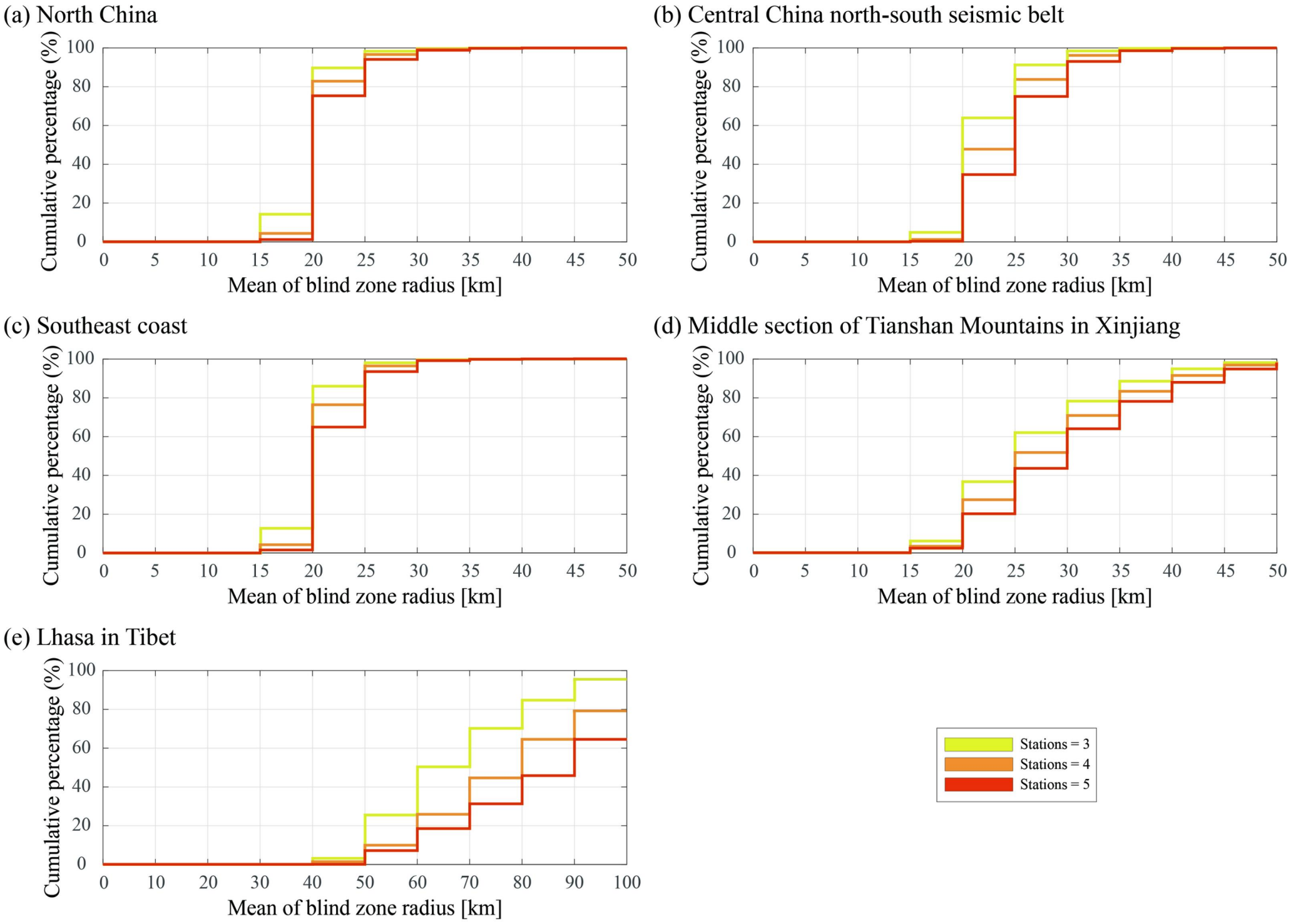

**Figure 5.** Cumulative distribution of mean blind zone radii in five key EEW regions (see Figure 1) for the cases of three, four, and five trigger stations, respectively.

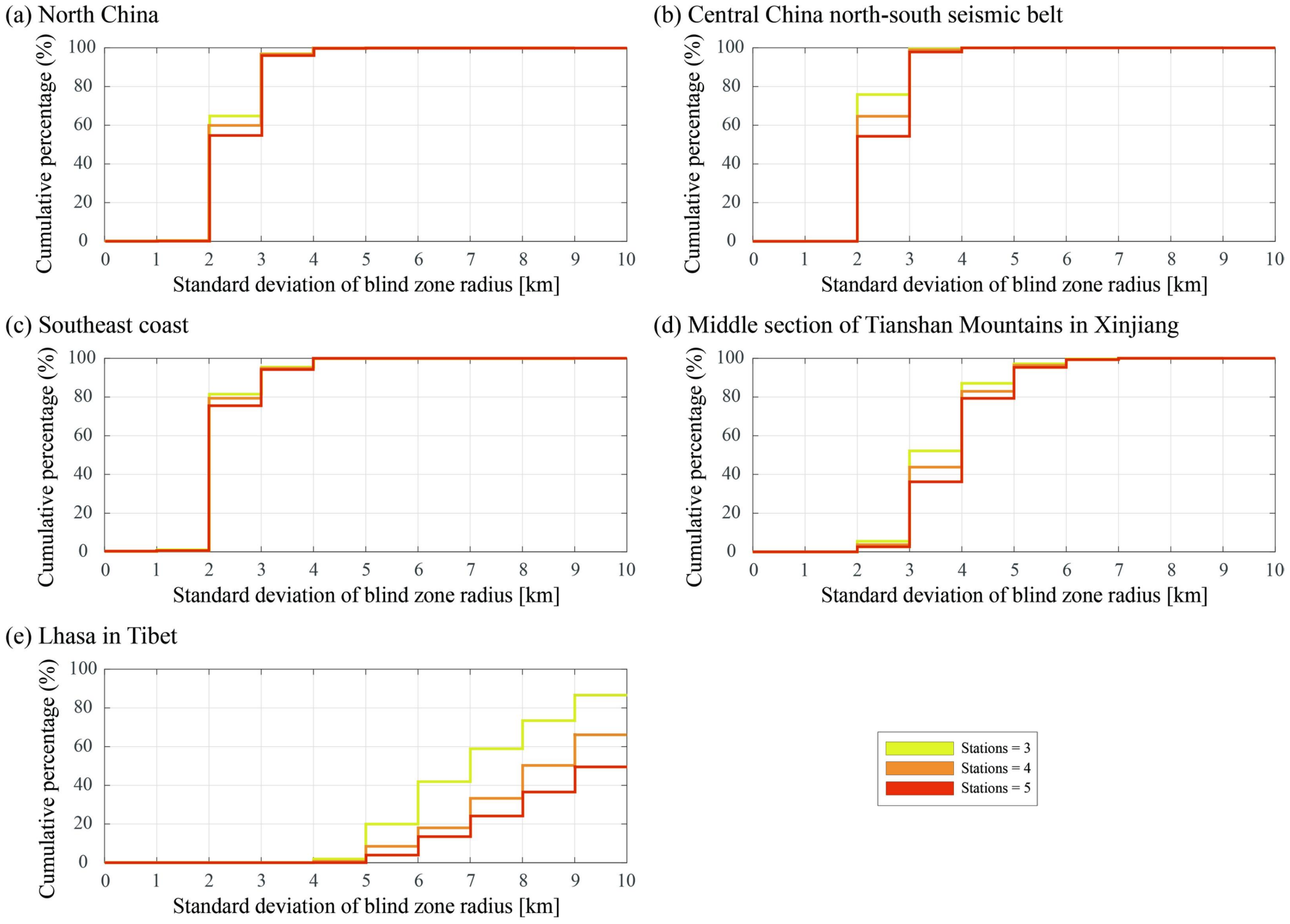

**Figure 6.** Cumulative distribution of standard deviation of the blind zone radii in five key EEW regions. Details are the same as Figure 5.

In addition, Figures 5 and 6 illustrate the impact of the "soft" blind zone on EEW performance in five regions for $i = 1$ and $i = 2$ failures of stations closest to the epicenter. In North China and the southeast coast, the areas with mean $r \leq 30$ km are reduced by factors of 0.95 to 0.98, and areas with mean $r \leq 20$ km decreased by 0.1 to 0.3 when the number of station failures increases from $i = 1$ to $i = 2$, compared to the case of $i = 0$ ($N + i = 3$). In the central China north-south seismic belt, the reduction factors range from 0.8 to 0.9 for mean $r \leq 30$ km and from 0.1 to 0.2 for mean $r \leq 20$ km. The middle section of the Tianshan Mountains in Xinjiang experiences greater reductions, ranging from 0.7 to 0.8 for mean $r$ smaller than 30 km and 0.4 to 0.5 for mean $r \leq 20$ km. In Lhasa, there is no area with mean $r \leq 50$ km in the presence of two station failures. In North China, the southeast coast and the central China north-south seismic belt, the areas with a standard deviation of $r$ smaller than 5 km do not change significantly when station failures occur. Therefore, it is crucial to continuously densify the network to control the "soft" blind zone for optimal functioning of the EEW system in the five regions.

## DISCUSSION

Globally acknowledged for their dense seismic monitoring networks and efficient EEW systems, regions like Japan, the U.S. West Coast, and Taiwan have set comparable standards. These regions have achieved approximately 30 km of comprehensive coverage, both for inter-station distances and blind zones (e.g., Kuyuk and Allen, 2013; Kohler et al., 2020). Similarly, the implementation of an advanced network in key EEW regions spanning North China, the central China north-south seismic belt, the southeast coast, and the middle section of the Tianshan Mountains in Xinjiang will result in a network density and coverage area with small-radius blind zones that can rival even the most extensive seismic networks worldwide, such as those in Japan, the U.S. west coast, and Taiwan. This enhanced network will offer a 30 km inter-station distance and blind zones in all key EEW regions, with the potential to cover key urban areas within a range of 10 to 20 km.

***The Cost-outcome Evaluation***

The costs per station of 1,000 new datum, 2,100 new basic, and 10,000 new ordinary stations of the planned network are 235,000 RMB, 130,000 RMB, and 10,000 RMB, respectively (Table 1; Qiang Ma, and Changsheng Jiang, written comm. 2021). Compared to the cost of building existing broadband seismic (350,000 RMB) and strong motion (200,000 RMB) stations in 2008 (Qiang Ma, and Changsheng Jiang, written comm. 2021), the station construction cost has decreased. This is mainly due to the localized development and production of seismometers, as well as the decline in their prices. The upgrade costs per station (to replace the data acquisition system and seismometer) for 1,000 existing broadband seismic and 1,100 strong motion stations are 125,000 RMB and 55,000 RMB, respectively. Therefore, the cost directly used to increase station density is 794 million RMB, which accounts for approximately 42% of the total investment (1.87 billion RMB).

Figures 1 and 2 show that the area covered by blind zones smaller than 30 km will increase from 1.7% to 22.3% inside Chinese mainland (11.5 million $km^2$), which means that this area will increase by 2.4 million $km^2$ (almost the size of Algeria). Assuming the improvement is made through the densification of the seismic network with an investment of 794 million RMB, we claim that every 1,000,000 RMB (146,000 USD in 2023) invested in the planned networks will increase the area where the blind zone radius is smaller than 30 km by 3,000 $km^2$. Similarly, every 1,000,000 RMB invested will increase the areas where the blind zone radius smaller than 20 km and 40 km by 250 $km^2$ and 4,300 $km^2$, respectively. The cost-outcome evaluation outlined above is admittedly rudimentary, as the computation hinges on a substantial assumption: a linear correlation between investment and the expansion of the area covered by a small-radius of blind zone. The credibility of this assumption naturally hinges on the specific placement of new seismic stations.

**Table 1.** Costs of the new stations in the planned seismic network in Chinese mainland.

| Items | DT station | BS station | OD station |
|---|---|---|---|
| The site construction | 90,000 | 60,000 | N/A |
| The data acquisition system | 50,000 | 40,000 | N/A |
| A broadband seismometer | 60,000 | N/A | N/A |
| An accelerometer | 15,000 | 15,000 | N/A |
| A MEMS sensor | N/A | N/A | 10,000 |
| other accessories | 20,000 | 15,000 | N/A |
| Total | 235,000 | 130,000 | 10,000 |

DT station: datum station; BS station: basic station; OD station: ordinary station. Costs are YUAN in RMB. Data from Qiang Ma and Changsheng Jiang (written comm. 2021).

***The System Alert Trigger Latency***

In this study, the theoretical prediction of $r$ for the planned network is based on the decomposition of the system alert trigger latency $T_{tri}$ in terms of $t_P$ and $T$. $t_P$ is determined by $\Delta$, $H$ and $V_P$, while $T$ is hypothesized to follow a normal distribution. Li et al. (2021) proposed the following empirical relationship for a line-source model-based EEW algorithm (Finite-Fault Rupture Detector algorithm, FinDer; Böse et al. 2012, 2015, 2018) in the Sichuan-Yunnan region, China:

$$T_{tri} = 2.5\exp(0.04D_{epi}) \quad (3)$$

where $T_{tri}$ is counted from the origin time of the earthquake with an inter-station distance $D_{epi}$ at its epicenter. The relationship is based on a playback investigation of 58 earthquakes ($5.0 \le M \le 8.0$). The algorithm uses a trigger threshold of 4.6 cm/s$^2$ and assumes a minimum number of three trigger stations. Additionally, it is assumed that there is no data transmission latency.

Equation (3) provides an alternative method for estimating $T_{tri}$. Previous studies have shown that $T_{tri}$ values obtained with other point-source-based algorithms differ slightly from those obtained with the FinDer algorithm (e.g., Li et al. 2020; Kohler et al. 2020). Therefore, we propose that Equation (3) can be adopted for EEW systems with the same trigger configurations to estimate the system alert trigger latency with the same uncertainty of $\sigma_{T\mathrm{tri}}$ = 1.8 s as determined by the regression conducted by Li et al. (2021). Figure 7 shows that the areas with empirical $T_{\mathrm{tri}}$ smaller than 5 s and 10 s account for 20% and 41%, respectively, which is close to the theoretical results for mean $r$ smaller than 30 km (22%) and 45 km (44%) in Figure 2a.

The findings of other research groups concerning the projected performance of the forthcoming EEW system in Chinese mainland align well with our predictions. Pan et al. (2019) anticipated a system alert trigger latency of 7 to 12 s in the southeastern region of Gansu, China, equating to a blind zone radius of 20 to 40 km (considering $V_{\mathrm{S}}$ = 3.57 km/s and $H$ = 15 km in their study). Our results in Figure 2a indicate a blind zone radius spanning 20 to 30 km for the same region, with an estimated uncertainty of 4 km. Peng et al. (2021) analyzed the EEW performance of the planned network during the earthquake sequence in Changning, Sichuan, between 2019 and 2020. They found that the system alert trigger latency $T_{\mathrm{tri}}$ was 6.3 ± 3.5 s (equivalent to a blind zone radius of 21 ± 13 km, assuming $V_{\mathrm{S}}$ = 3.5 km/s and $H$ = 7 km). This finding aligns reasonably well with our predicted values of $T_{\mathrm{tri}}$ (4 ± 1.8 s) in Figure 7 and the blind zone radii $r$ (21 ± 3 km) in Figure 2a for the same region. However, there was a slightly larger deviation in the results reported by Peng et al. (2021), likely stemming from the complexities of real-world conditions that extend beyond the simplifications we made in our parameter assumptions.

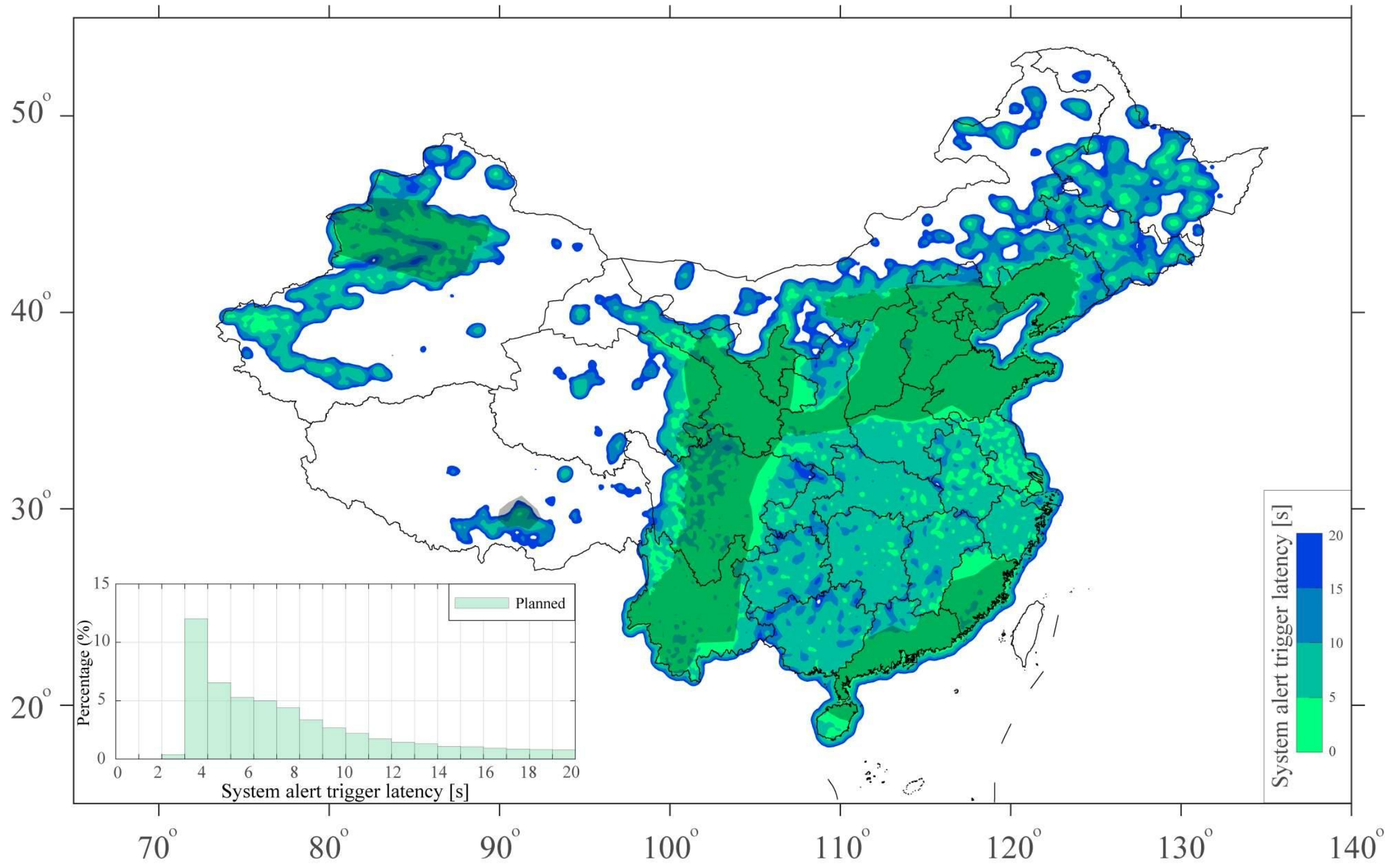

**Figure 7.** Maps showing the estimated system alert trigger latency $T_{\mathrm{tri}}$ from the earthquake origin time for the planned EEW system. Areas where $T_{\mathrm{tri}}$ surpass 20 s are displayed in white. The $T_{\mathrm{tri}}$ was calculated using Equation (3). The histogram in the bottom left shows the binned distribution of $T_{\mathrm{tri}}$ inside Chinese mainland. Details are the same as Figure 1.

The current test of the planned EEW system in the Chinese mainland implements a relatively complex set of system alert trigger criteria to mitigate false alerts (Wenhui Huang, written comm. 2023). Firstly, different trigger thresholds have been set for each type of station based on their data observation quality, starting with datum stations having the best quality, followed by basic stations, and lastly ordinary stations with the worst quality. Secondly, the system applies a double-checking mechanism to ensure that the alert triggers are not caused by non-earthquake events. This is done by analyzing the signal-to-noise ratio of the data, the waveform duration of events, and the spatial configuration of trigger and non-trigger stations in real-time. Finally, a system alert is issued only after confirming that an earthquake occurred. Despite the complexity of this procedure, the system latency $T$ is reasonably small based on practical test runs (Wenhui Huang, written comm. 2023). Therefore, $T$ is modeled to follow a normal distribution $N(4, 0.5^2)$ for the planned network in this study.

However, it is worth noting that the system alert trigger criteria are not directly used or considered in our calculation. Rather, their significance is indirectly reflected in the simulation of $T$. If the current alert trigger criteria remain unchanged, our simulation for $T$, and therefore estimation of $r$, is reasonable. On the one hand, our estimation of the blind zone size based on the simulation of $T$ and a minimum number of three trigger stations is conservative. Typically, the system would trigger an alert once two stations are triggered, and source parameters can be estimated using the early part of the event waveform in the test run of the planned network. On the other hand, our estimation of the blind zone based only on $T$ may be optimistic and correspond to a lower bound. Indeed, the size of the blind zone could be larger than our theoretical calculation if additional latency is considered, such as communication and response latency needed to take efficient actions. It is important to note that the blind zone evaluated in our study is determined from the properties of the EEW system rather than from considerations involving earthquake hazard risk, which should consider additional factors related to the earthquake impact on society. Whatever the end use, the blind zones determined from the EEW system is useful in guiding the optimization of system construction and earthquake hazard risk reduction strategies.

## CONCLUSIONS

By simply densifying the seismic network layout, the performance of an EEW system can be significantly improved. As of 2023, approximately 80% of the area inside Chinese mainland will have inter-station distances of less than 100 km, with 50%, 30%, and 17% having inter-station distances of 50 km, 25 km, and 15km, respectively. These proportion of inter-station distances are expected to expand by factors of 1.4, 2.2, 4.4, and 10 in the near future, compared to that of the existing layout. In this study, we predicted the spatial distribution of the blind zone radii and quantified the warning effectiveness of the planned EEW system in Chinese mainland based on its network configuration. As a novelty, we employed Monte Carlo random sampling with 50,000 iterations to account for uncertainty in the estimate of $r$. Additionally, we investigated the impact of "soft" blind zones caused by potential station failures on the performance of the planned EEW system in China.

We find that $r$ is inversely correlated only with $H$ and $V_S$, and $H$ contributes significantly to the occurrence of zero blind zones. Taking into account the practical change amplitude scale of parameter related to $r$ in real-world, $\Delta$ and $H$ have the greatest impact on $r$, resulting in a magnitude change of $r$ of $10^1$ km and $-10^1$ km. The area covered by blind zones smaller than 30 km will increase by a factor of 13 from 1.7% to 22% inside Chinese mainland, which means that this area will increase by 2.4 million km$^2$ (almost the size of Algeria). The planned network is expected to achieve almost complete coverage of areas with mean $r \leq 30$ km in North China (98%), the southeast coast (98%), and the central China north-south seismic belt (91%). However, the middle section of the Tianshan Mountains in Xinjiang (63%) and Lhasa in Tibet exhibit weaker EEW performance, and achieving effective EEW for Lhasa in Tibet is difficult. Additionally, areas with standard deviation of $r$ less than 5 km and 10 km will cover 46% and 77% of the areas inside Chinese mainland, respectively. This implies that densification of the networks will effectively constrain the potential variability in the size of the blind zones, which is significant for the successful functioning of the nationwide EEW system. Considering that stations located near the epicenter may fail to function correctly or underestimate the hazards of ground-shaking, the proportion of areas with mean $r \leq 30$ km and 40 km can decrease by a factor of 0.84 to 0.85 for one station failure and by 0.72 to 0.75 for two station failures. To prevent the unforeseen enlargement of blind zones due to common station failures, it is still necessary to continue to densify the seismic network in some key regions where the blind zone radii range from 15 to 40 km.

Assuming that the above improvement is made through the densification of the seismic network, we claimed that every 1,000,000 RMB (146,000 USD in 2023) invested in the planned networks will increase the area where the blind zone radius is smaller than 30 km by 3,000 km$^2$. Similarly, every 1,000,000 RMB invested will increase the areas with $r \leq 20$ km and 40 km by 250 km$^2$ and 4,300 km$^2$, respectively.

Our work not only offers insights into the anticipated performance of the forthcoming EEW network in Chinese mainland but also introduces a broadly applicable evaluation approach for predicting the performance of any EEW system during its planning, design, and implementation stages. This investigation serves as a valuable reference for the further enhancement and optimization of the EEW system in Chinese mainland. Further quantitative exploration of how expanding coverage area with smaller-radius of blind zones (e.g., 30 km) could effectively mitigate earthquake risks by reducing exposure to population, the economy, essential lifelines, and infrastructure warrants consideration.

## DATA AND RESOURCES

The earthquake catalog and information for the existing and the planned seismic networks were provided by the China Earthquake Networks Center (doi: 10.11998/SeisDmc/SN) through the internal link of the Earthquake Cataloging System at China Earthquake Administration (http://10.5.160.18/console/index.action. The total investment information for the National System for Fast Report of Intensities and Earthquake Early Warning Project of China was obtained from the China Earthquake Networks Center (https://www.cenc.ac.cn/cenc/zt/361404/361414/361563/index.html, last accessed in January 2022). The fifth-generation earthquake hazard zoning map of China can be accessed through the link: https://www.gb18306.net/ (last accessed in March 2023).

## Statements and Declarations

This work was supported by the Guangdong Basic and Applied Basic Research Foundation (Grant numbers 2020A1515110844), the Special Fund of the Institute of Geophysics, China Earthquake

Administration (Grant numbers DQJB22Z01-10), and the National Natural Science Foundation of China (Grant numbers U2039202 and U2039207). The authors have no relevant financial or non-financial interests to disclose. All authors contributed to the study conception and design. Material preparation, data collection and analysis were performed by Jiawei Li. The first draft of the manuscript was written by Jiawei Li and all authors commented on previous versions of the manuscript. All authors read and approved the final manuscript.

**Estimating Earthquake Early Warning Effectiveness via Blind Zone Sizes: A Case Study of the Planned Seismic Network in Chinese Mainland**

Jiawei Li[1)], Didier Sornette[1)], and Yu Feng[1,2)]

1. Institute of Risk Analysis, Prediction and Management (Risks-X), Academy for Advanced Interdisciplinary Studies, Southern University of Science and Technology (SUSTech), Shenzhen, China, 518055

2. Norwegian Geotechnical Institute, Oslo, Norway, N-0484

(Corresponding author: Jiawei Li; lijw@cea-igp.ac.cn; lijw3@sustech.edu.cn)

Using a theoretical network-based method along with Monte Carlo random sampling with 50,000 iterations, we predicted the spatial distribution of the blind zone radii and quantified the warning effectiveness of the planned EEW system in Chinese mainland based on its network configuration. Additionally, we investigated the impact of “soft” blind zones caused by potential station failures on the performance of the planned EEW system in China. This file serves as a valuable complement to the main text, featuring earthquake plots, existing and planned station distributions across Chinese mainland, as well as maps illustrating inter-station distances for both the existing and planned networks. Additionally, included are maps indicating alterations in mean blind zone radii between the two networks, requiring a minimum of three trigger stations. Furthermore, estimates of mean and standard deviation of blind zone radii with a minimum of four and five trigger stations are also presented here.

**Contents of this file**

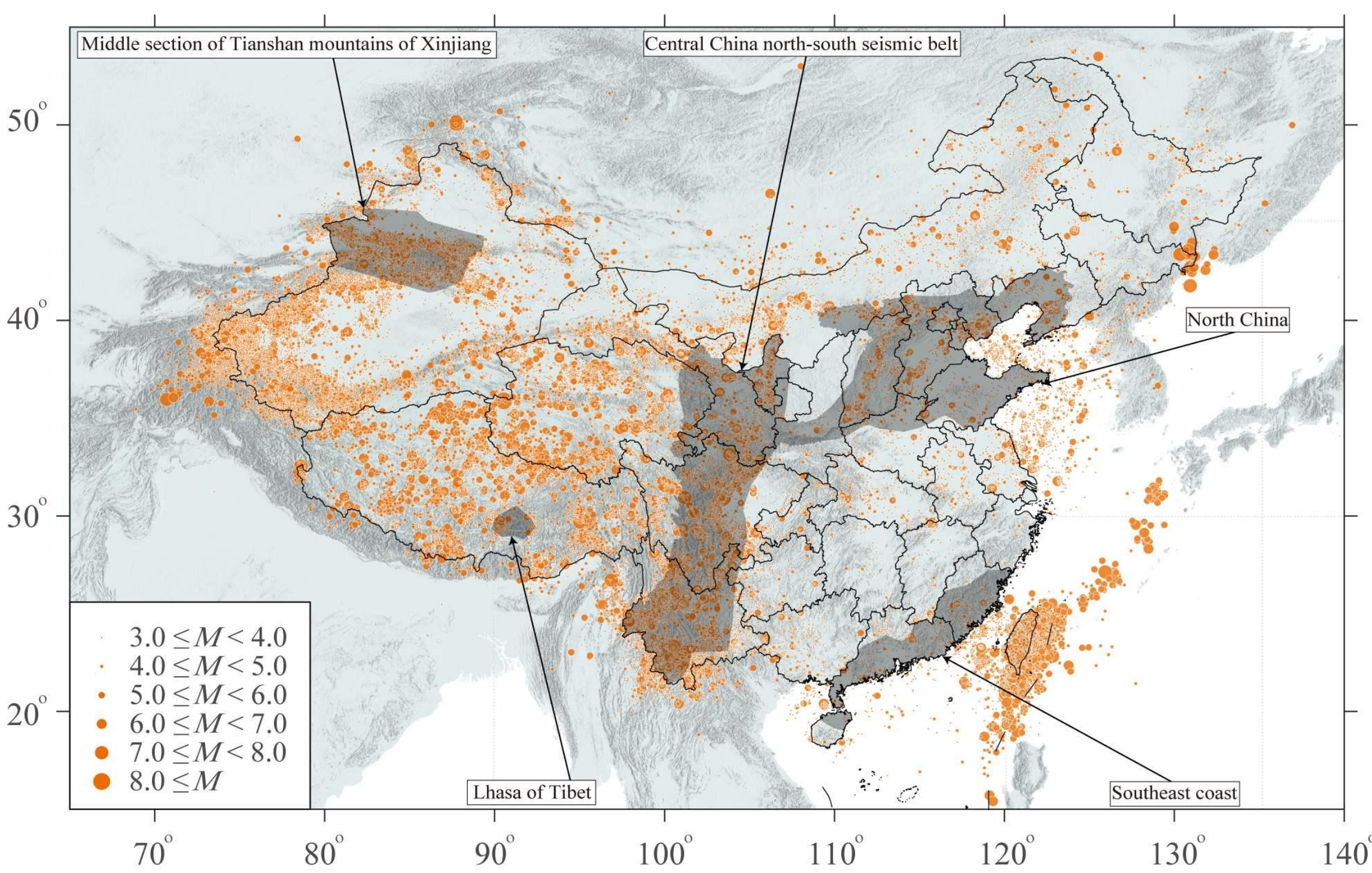

**Figure S1.** The epicenters (from China Earthquake Networks Center, CENC) of 91,981 $M_L \geq 3.0$ earthquakes from 01/01/1970 to 29/06/2022 in China and its adjacent regions. The darker areas represent the five key earthquake early warning (EEW) regions that were developed by the National System for Fast Report of Intensities and Earthquake Early Warning Project of China and are expected to be operational after 2023.

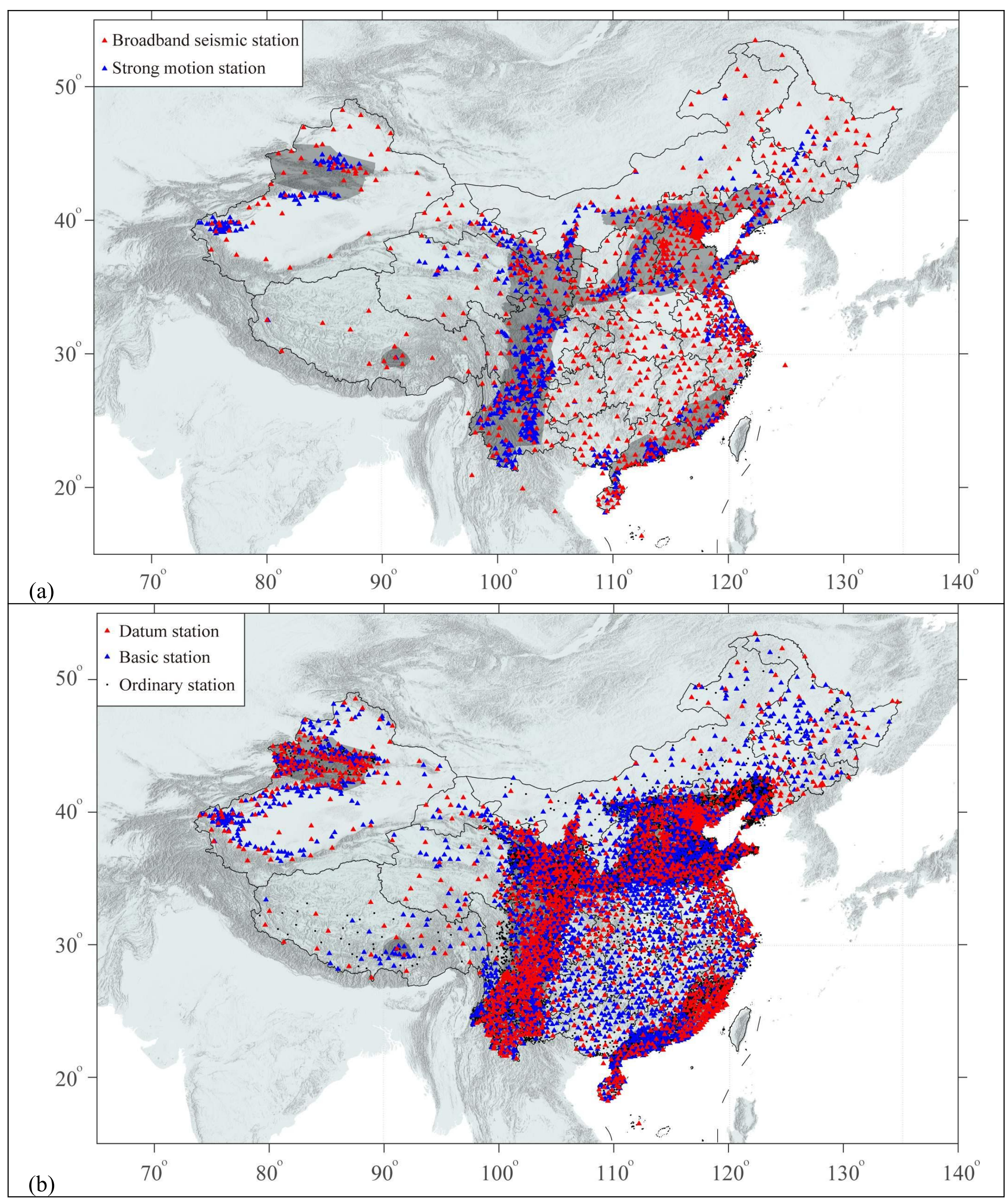

**Figure S2.** (a) Existing (as of 2018) stations as part of the CENC and the China Strong Motion Network Center (CSMNC). (b) Planned (as of 2023) stations to be newly deployed or upgraded in the National System for Fast Report of Intensities and Earthquake Early Warning of China. The darker areas represent the five key earthquake early warning (EEW) regions.

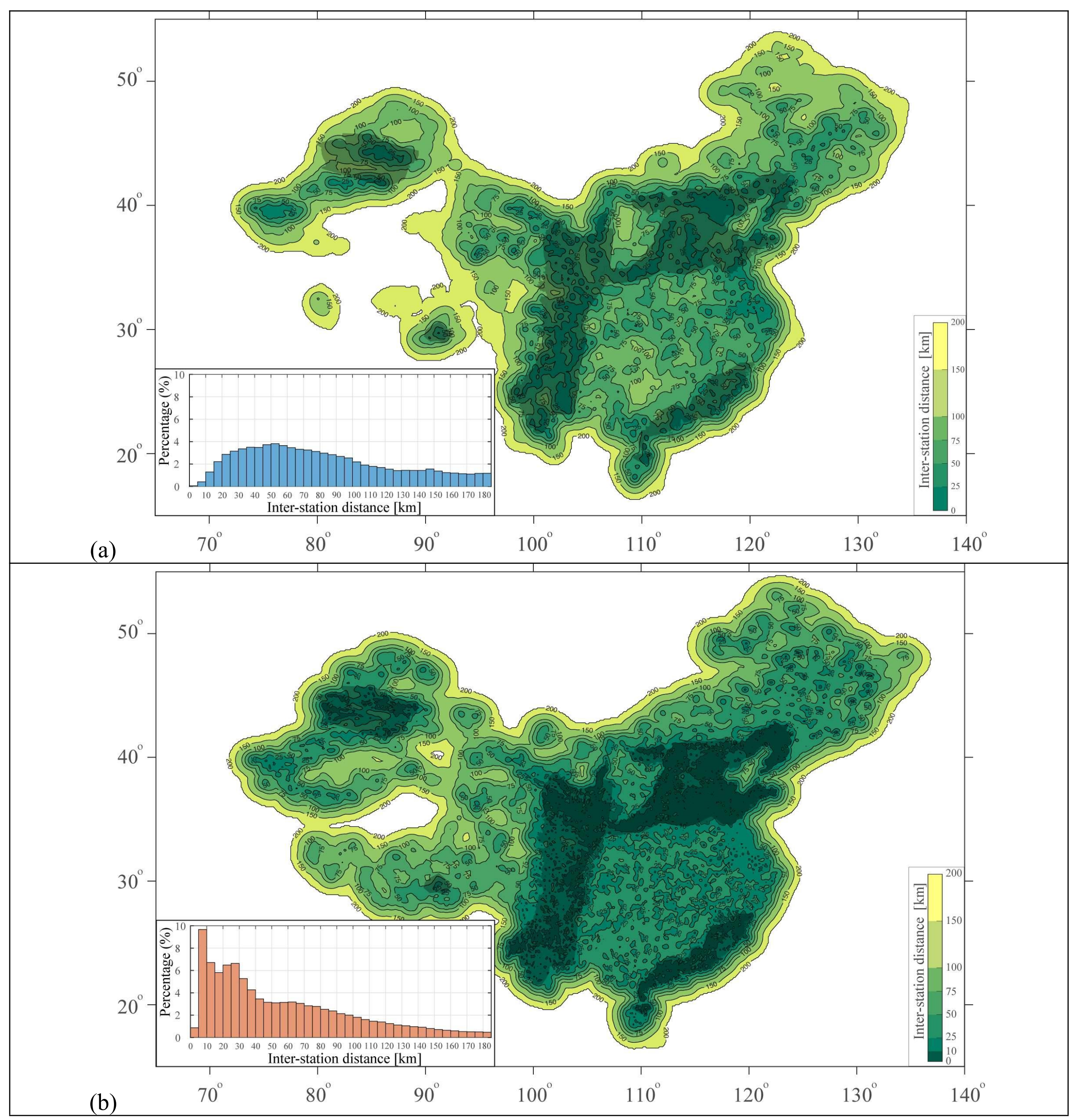

**Figure S3.** Maps showing the inter-station distances for the (a) existing and (b) planned networks. For any given site (assuming a grid of 0.1° × 0.1° resolution), the inter-station distance have been calculated as the average distance to the three closest stations (Kuyuk and Allen 2013; Li et al. 2016; Li et al. 2021). The histograms in the bottom left depict the binned distribution of the inter-station distances inside Chinese mainland. We define "inside" in this study as the region within 100 km outside of Chinese mainland's boundaries.

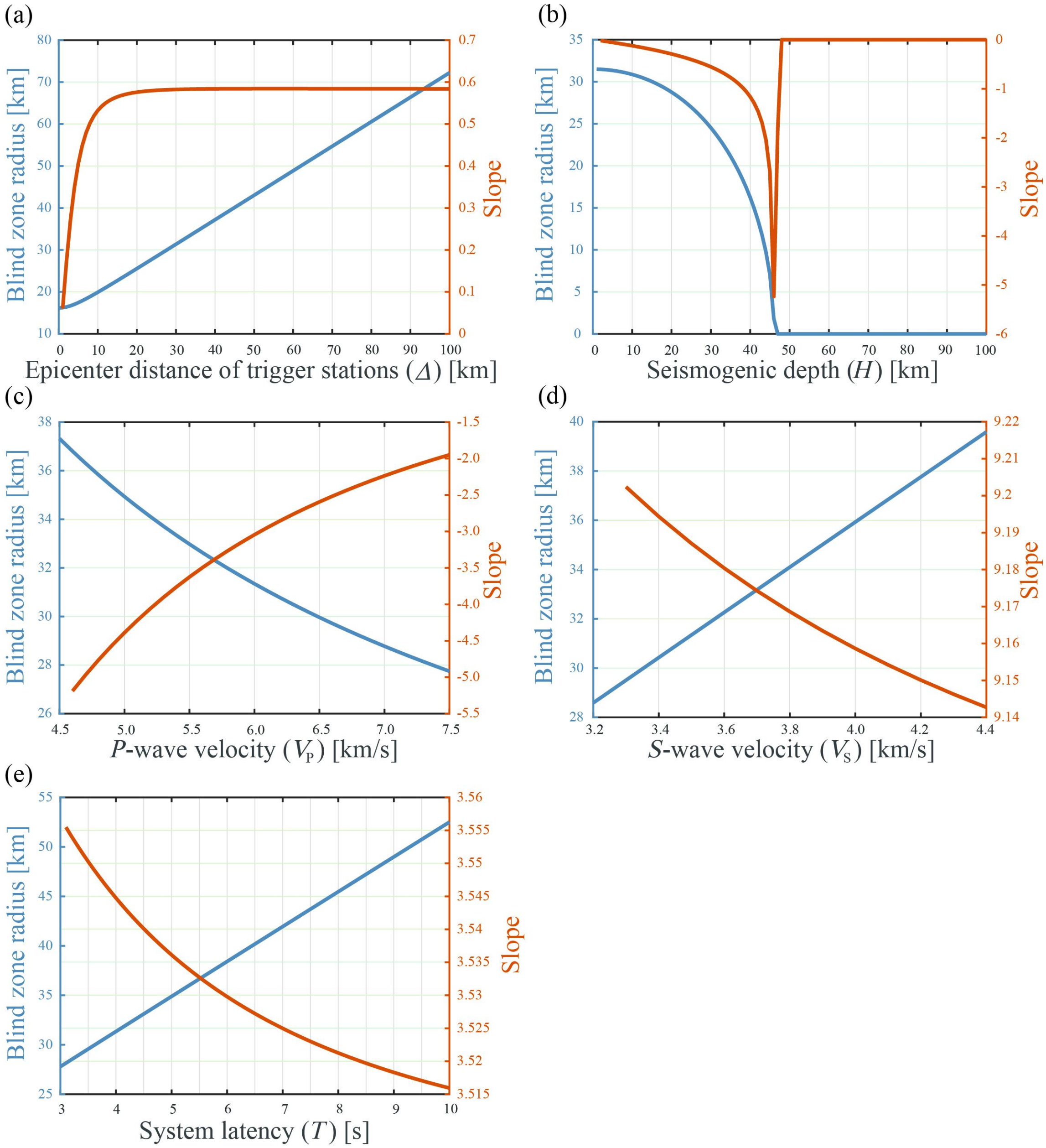

**Figure S4.** Dependence of blind zone radius (left axis) on various parameters: (a) epicenter distance ($\Delta$) of trigger stations, (b) seismogenic depth ($H$), (c) $P$-wave velocity ($V_P$), (d) $S$-wave velocity ($V_S$), and (e) system latency ($T$). The blind zone radius has been calculated by sequentially varying each parameter while keeping the remaining four parameters fixed at constant values of $\Delta$ = 30 km, $H$ = 5 km, $V_P$ = 6 km/s, $V_S$ = 3.5 km/s, and $T$ = 4 s. The slope of the curve corresponding to each variable is represented on the right axis in each panel.

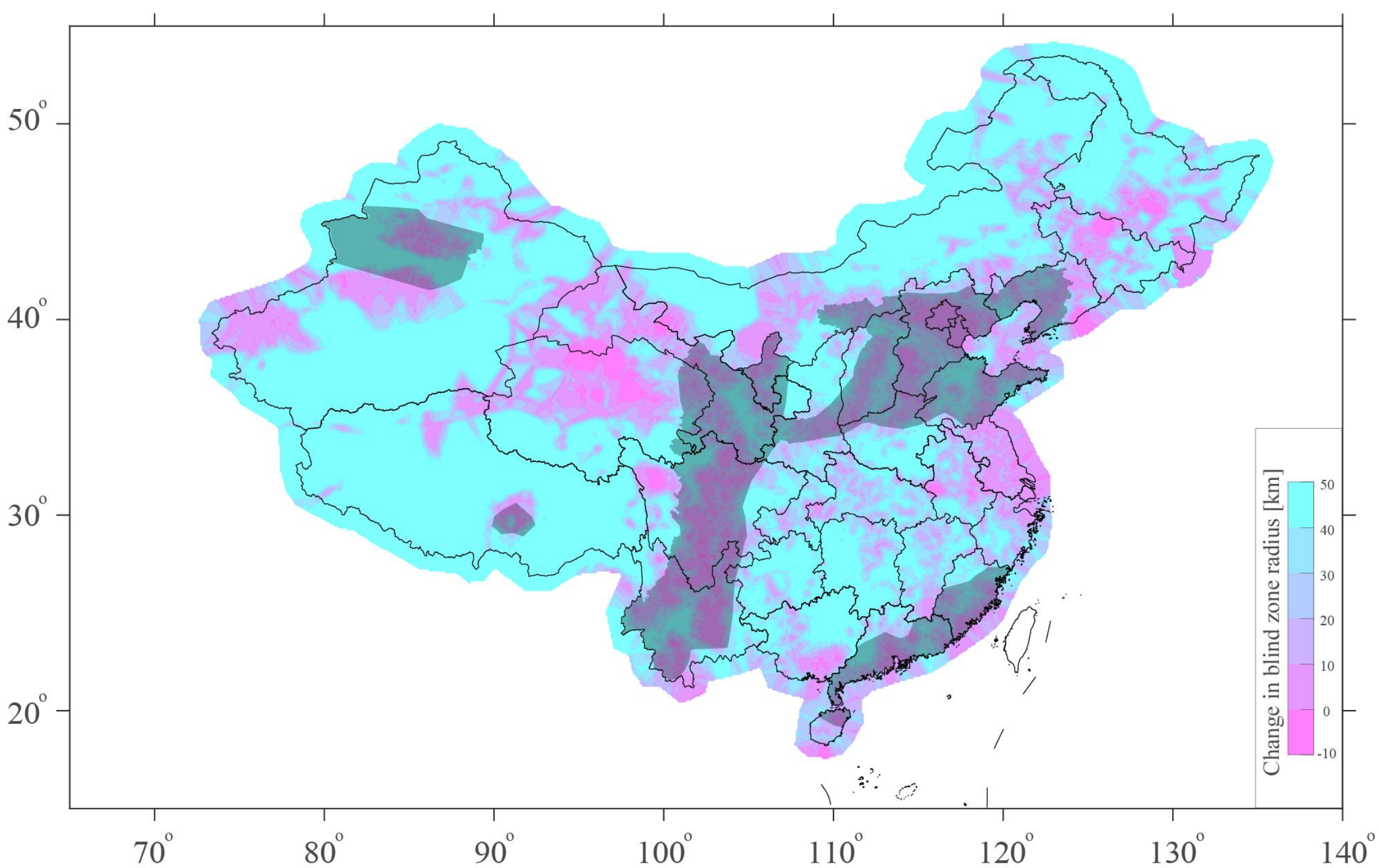

**Figure S5.** Map showing changes in mean blind zone radii between the existing and planned networks with a minimum of three trigger stations (Figures 2a and 2b). Positive values indicate relative reduction in mean blind zone radii of the planned network compared to the existing network, while negative values indicate expansion of mean blind zone radii of the planned network relative to the existing network. Limited to inside regions of Chinese mainland, encompassing both comprehensive coverage of the existing and planned networks.

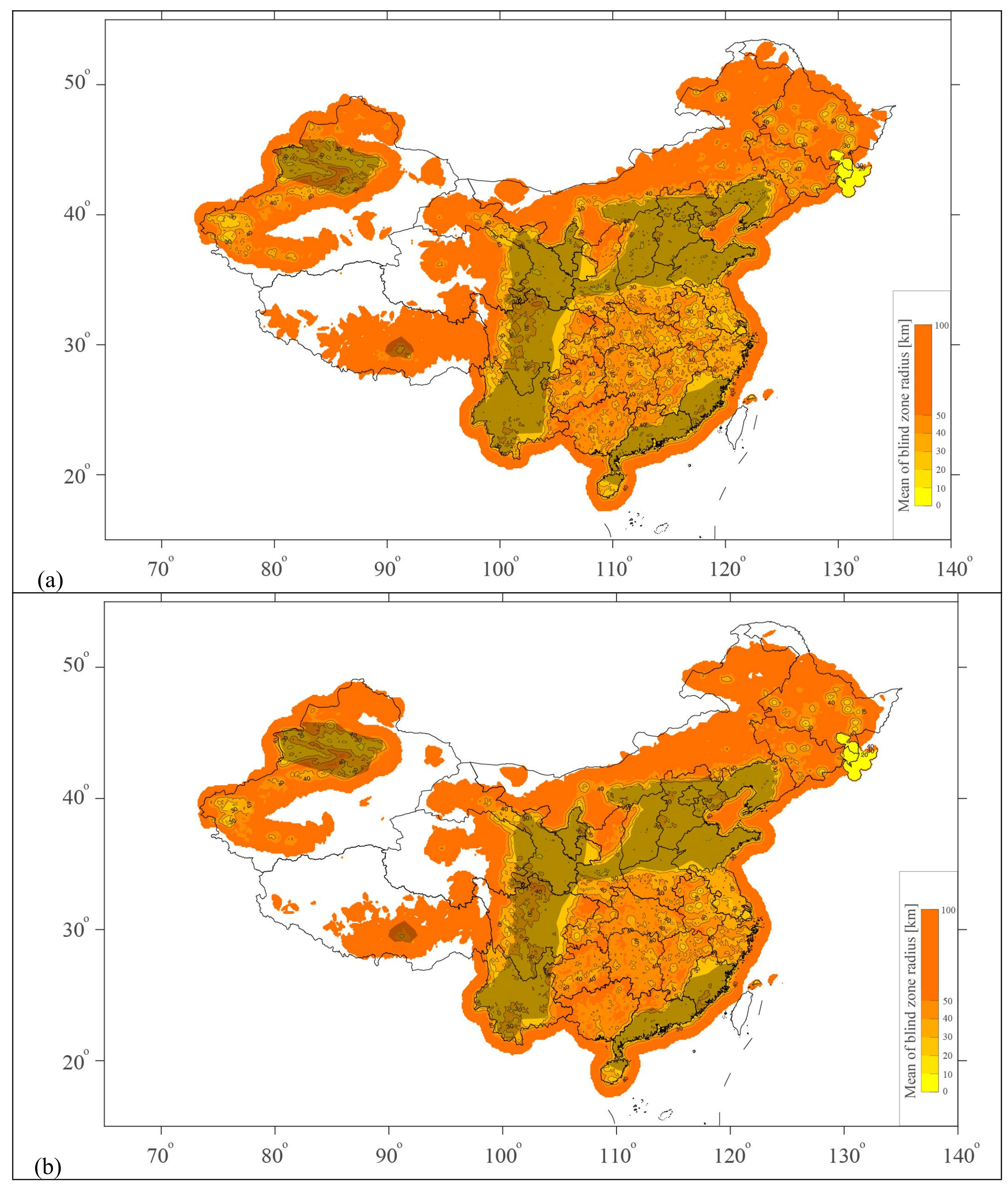

**Figure S6.** Maps showing the mean of the estimated blind zone radii for the planned network assuming a minimum number of (a) four and (b) five trigger stations, respectively. Details are the same as Figures 1a and 2a. The curves of the cumulative distribution inside Chinese mainland can be found in Figure 4a.

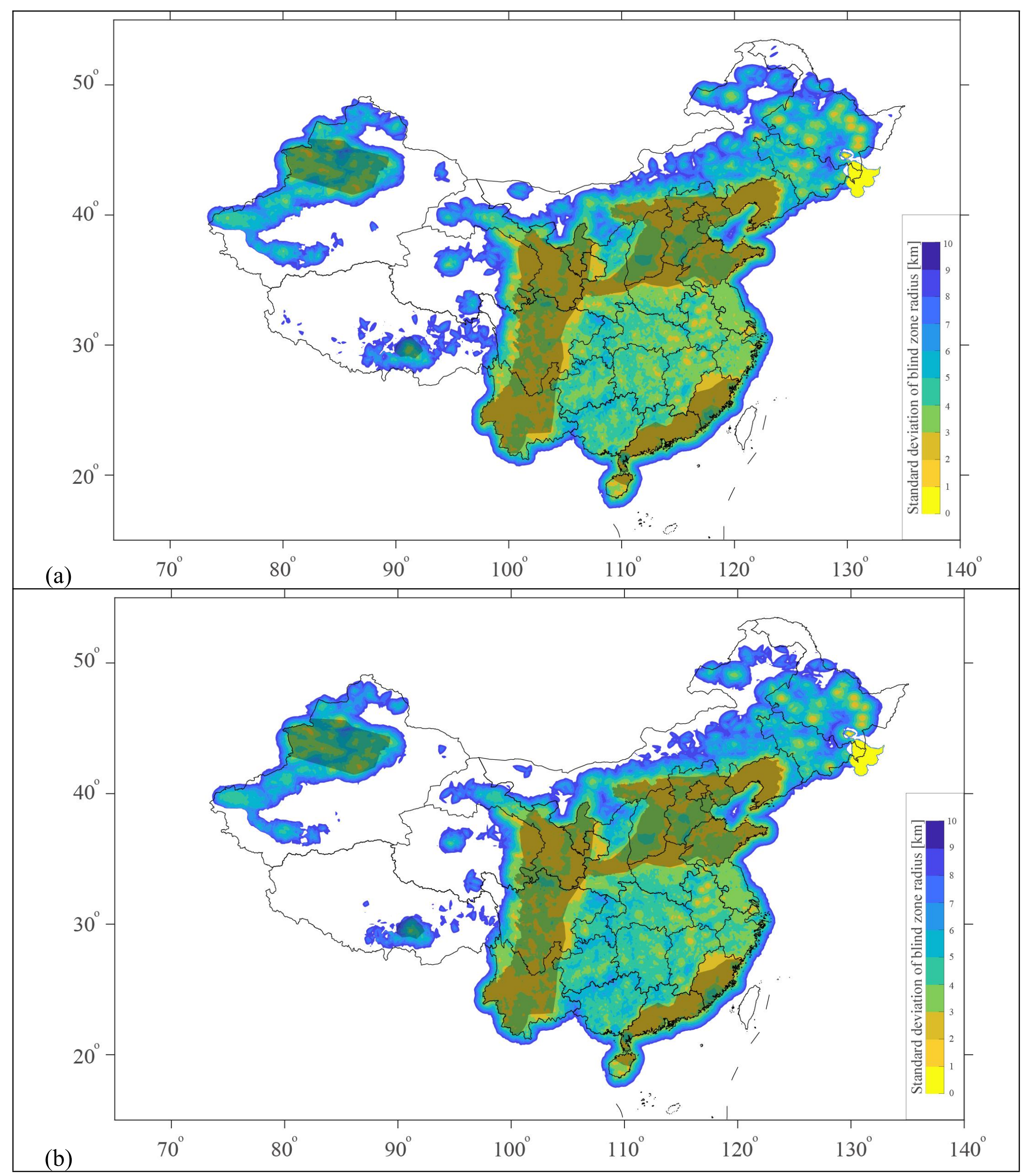

**Figure S7**. Maps showing the standard deviation of the estimated blind zone radii for the planned network assuming a minimum number of (a) four and (b) five trigger stations, respectively. Details are the same as Figures 1b and 2b. The curves of the cumulative distribution inside Chinese mainland can be found in Figure 4b.